# Advanced Nanostructured Topical Therapeutics for Psoriasis: Strategic Synthesis, Multimodal Characterization, and Preliminary Pharmacodynamic Profiling

Iqra Yousaf[1] and Aqsa Yousaf[2]

[1]Department of Chemistry, University of Engineering and Technology, Lahore, Pakistan, 54890
[2]Department of Computer Science and Engineering, University of Texas at Arlington, USA

**Abstract**

Psoriasis is a chronic, immune mediated inflammatory skin condition that poses substantial therapeutic challenges. In this study, we developed a novel topical nanotherapeutic system integrating metal oxide nanoparticles and botanical antioxidants into a fish collagen and agar gel base. Cerium oxide ($CeO_2$), zinc oxide (ZnO), and silver (Ag) nanoparticles were synthesized and characterized using UV–Vis spectroscopy, dynamic light scattering (DLS), FTIR, and scanning electron microscopy (SEM). The nanoparticles demonstrated stable dispersion, narrow size distribution (e.g., ZnO averaged 66 nm), and strong interactions with the gel matrix. To enhance therapeutic potential, the gel was enriched with plant based antioxidants from *Momordica charantia* (bitter melon), *Zingiber officinale* (ginger), and *Azadirachta indica* (neem). The resulting multicomponent formulation was evaluated in a psoriasis induced animal model.Lesion size measurements and statistical analyses revealed significantly accelerated wound contraction and reduced inflammation in the treatment group compared to placebo and untreated controls ($p < 0.01$ to $p < 0.001$) across multiple time points). These effects were observed as early as Day 3 and became more pronounced by Day 14. Our findings highlight the synergistic anti-inflammatory, antioxidant, and antimicrobial actions of the nanoparticle enriched botanical gel. This approach addresses key aspects of psoriasis pathophysiology, offering a promising therapeutic avenue. Further validation in clinical and psoriasis specific models is recommended to optimize formulation and evaluate long term safety.

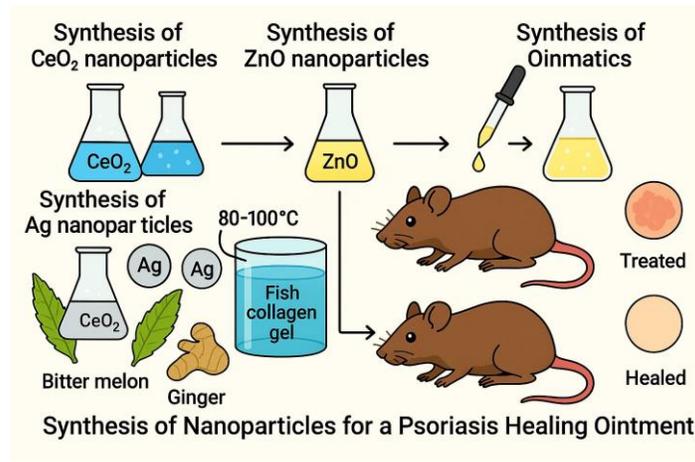

**Synthesis of CeO₂ nanoparticles** — **Synthesis of ZnO nanoparticles** — **Synthesis of Oinmatics**

Synthesis of Nanoparticles for a Psoriasis Healing Ointment

# 1   Introduction

## 1.1   Overview of Psoriasis

Psoriasis is a persistent and challenging inflammatory condition that affects approximately 2–3% of the global population (Grozdev et al., 2014). While it is most commonly recognized by its recurring skin symptoms, psoriasis is increasingly viewed as a systemic immune-mediated disorder. It shares pathophysiological features with other chronic inflammatory diseases such as rheumatoid arthritis and Crohn's disease conditions rooted in immune system dysregulation and often managed using similar therapeutic approaches (Mrowietz and Reich, 2009). This growing body of evidence supports psoriasis's classification among immune mediated inflammatory diseases (IMIDs), highlighting its deeper immunological implications beyond skin involvement.

Epidemiological studies reveal a higher prevalence of various noncutaneous comorbidities in individuals with psoriasis (Henseler and Christophers, 1995). These comorbidities are typically divided into two groups: the first includes those with a direct pathogenic relationship to psoriasis, such as psoriatic arthritis, Crohn's disease, and pustular dermatoses; the second group includes conditions like metabolic syndrome and cardiovascular diseases, which are linked indirectly due to the long-standing inflammatory burden of the disease (Sanz, 2007). These secondary conditions often emerge over time and may affect multiple organ systems. Importantly, while they are considered secondary, their cumulative impact on patient health can be more substantial than the primary dermatologic symptoms (Grozdev et al., 2014).

## 1.2    Comorbidities and Systemic Impact

Psoriasis is frequently associated with a variety of comorbid conditions that reflect its systemic nature (Daugaard et al., 2022). Epidemiological studies show that people with psoriasis have higher rates of other chronic illnesses compared to the general population (Daugaard et al., 2022). These comorbidities can be broadly grouped into two categories.

**Immune Related Conditions:** Diseases that share direct pathogenic links with psoriasis. Examples include psoriatic arthritis (an inflammatory arthritis occurring in up to 30% of psoriasis patients) and Crohn's disease or other inflammatory bowel diseases. These conditions are thought to arise from similar immune system dysregulation that drives psoriasis.

**Cardiometabolic Conditions:** Disorders that are associated with the chronic inflammatory burden of psoriasis over time. This group includes metabolic syndrome (obesity, insulin resistance, dyslipidemia, hypertension) and cardiovascular diseases such as heart disease and stroke. Long term systemic inflammation in psoriasis is believed to contribute to these risk factors.

Researchers propose that many of these co occurring conditions are secondary manifestations of psoriasis's systemic inflammation, emerging over the course of the disease and affecting multiple organ systems. Importantly, the impact of psoriasis associated comorbidities on a patient's health can be profound – in some cases posing greater risks than the skin lesions themselves. For instance, severe psoriasis has been linked to increased cardiovascular events and mortality, underscoring the need to manage psoriasis as a multi-system disease.

## 1.3    Limitations of Current Treatments

Multiple therapeutic options exist to treat psoriasis, including topical creams, phototherapy, systemic immunosuppressants, and biologic drugs that target specific immune pathways. These conventional treatments can substantially improve skin lesions and patient quality of life. These conventional treatments can substantially improve skin lesions and patient quality of life (Lee and Kim, 2023). However, psoriasis remains a lifelong condition with no definitive cure and many patients encounter challenges with current therapies.

**Adverse Effects and Safety:** Potent systemic treatments (like oral retinoids, methotrexate, or biologics) can cause significant side effects and raise long term safety concerns. Patients may experience anything from skin irritation (with topicals) to organ toxicity or infection risk (with immunosuppressants).

**Variable Efficacy and Tolerance:** Individuals respond differently to psoriasis therapies. Some fail to achieve adequate symptom control or lose responsiveness over time (treatment resistance). It often requires trial and error to find an effective regimen, and even then, relapses can occur.

**High Costs:** Advanced treatments such as biologic injections are expensive. The high cost and need for regular dosing can limit patient access or adherence to therapy, especially for those without sufficient insurance coverage. **Chronic Management Burden:** Because psoriasis is chronic and tends to recur, patients must use treatments continuously or in repeated courses. This long term management can strain patients financially and psychologically, and maintaining adherence is difficult when balancing efficacy and side effects over years of therapy.

These limitations in efficacy, safety, and practicality have spurred interest in novel therapeutic approaches. There is a growing need for treatments that can more selectively target psoriatic disease mechanisms with fewer side effects and more durable results. One emerging strategy is to leverage nanotechnology to improve drug delivery and therapeutic precision in psoriasis management.

### 1.4    Nanotechnology in Psoriasis Treatment

Significant global interest has emerged in various aspects of nanotechnology research, resulting in numerous advancements and novel applications of diverse nanoparticles in fields such as energy, electronics, space exploration, and medicine (Abdulsattar, 2015). Nanoparticles are defined by their size, distribution, and morphology, typically comprising clusters of atoms ranging from 1 to 100 nm. Due to their nanoscale dimensions, these particles possess unique and enhanced properties compared to their larger counterparts derived from bulk materials ((Saliem et al., 2016).

## 2    Nanoparticles in Dermatological Applications

### 2.1    Zinc Oxide Nanoparticles in Dermatology

For many years, zinc oxide (ZnO) has been widely used in food, pharmaceuticals, cosmetics, and other chemicals used daily, and formulations of these products containing ZnO nanoparticles (ZnO NPs) are more popular than conventional formulations (Aditya et al., 2018). Topical medications and cosmetics that come into frequent contact with the skin often contain a high concentration of ZnO nanoparticles (Osmond-McLeod et al., 2014). The topical use of ZnO nanoparticles (NPs) on healthy human skin is generally considered safe. However, individuals with skin conditions such as psoriasis, atopic dermatitis (AD), acne, and rosacea often use ZnO NP based medicinal or cosmetic products as frequently as, or even more frequently than, those with normal skin. These products are commonly utilized for treatment purposes or to conceal visible skin imperfections. Despite their widespread use, there is limited discussion regarding the safety of ZnO NPs on diseased skin, and existing guidelines do not offer clear recommendations on the matter (Lai et al., 2021).Inflammatory skin diseases are often characterized by impairments in the skin barrier and irregular expression of barrier associated proteins in the epidermis.For instance, both atopic dermatitis (AD) and psoriasis result in the dysregulated expression of filaggrin and loricrin, which are essential for maintaining the structural integrity of the epidermal barrier (Thyssen et al., 2020). In a mouse model of atopic dermatitis (AD), ZnO NPs have been reported to penetrate the dermis through the epidermis, triggering an intense IgE response (Ilves et al., 2014). Consequently, the skin affected by acne, rosacea, and psoriasis is likely to be permeable to ZnO nanoparticles, increasing the likelihood of their penetration. This suggests that keratinocytes in the stratum basale are at a heightened risk of exposure to ZnO NPs. Additionally, the majority of studies indicate that ZnO nanoparticles exert proinflammatory and cytotoxic effects on keratinocytes in vitro, with oxidative stress being identified as the primary mechanism responsible for cellular damage.

## 2.2 Silver Nanoparticles for Anti-Psoriatic Therapy

Silver nanoparticles (Ag NPs) are among the most significant and appealing metallic nanoparticles, particularly due to their relevance in biomedical applications (Keck and Schwabe, 2009), (Samberg et al., 2010), (Zhang et al., 2016). Ag NPs are extensively utilized in nanomedicine, particularly for cancer diagnosis and treatment. However, their potential application in antimicrobial delivery remains a significant area of interest (Jung et al., 2009). Ag NPs have been effectively utilized as carriers for delivering anti-psoriatic drugs. The nanoparticles ranged in size from 20 to 80 nm. Their anti-inflammatory potential was assessed in vitro using UVB exposed HaCaT cells, while their in vivo efficacy was evaluated on human psoriatic lesions and an acute inflammation model. The advanced AgNP formulation demonstrated significant anti-inflammatory properties in vitro. Additionally, reduced edema and decreased cytokine levels were observed in paw tissue. Ultimately, Ag NPs were identified as a promising vehicle for the targeted delivery of anti-psoriatic therapeutics.

## 2.3 Role of Cerium Oxide Nanoparticles in Oxidative Stress Management

Cerium oxide nanoparticles generally range in size from 1 to 100 nm, providing a high surface area to volume ratio. This characteristic enhances their surface reactivity and significantly influences their effectiveness as catalysts (Baldim et al., 2020). In psoriasis, excessive inflammation often goes hand in hand with oxidative tissue damage, which can slow down recovery of the skin.

Oxidative stress is characterized by an imbalance between free radicals and antioxidants generated during the wound healing process, which has been shown to hinder and slow the progression of healing (Deng et al., 2021). The high surface reactivity boosts the ability to interact with and neutralize multiple ROS in the cellular environment. By reducing oxidative stress, $CeO_2$ NPs indirectly mitigate inflammation and tissue damage in the skin. In wound healing studies, cerium oxide nanoparticles have demonstrated remarkable therapeutic effects: they accelerate the closure of skin wounds, reduce inflammation and scar formation, and even exhibit antibacterial activity to prevent infection (Chen et al., 2024).

$CeO_2$ NPs exhibit potent antioxidant properties by replicating the activity of key enzymes in the body, including superoxide dismutase and catalase (Saifi et al., 2021). CeNPs maintain a dynamic equilibrium between Ce (III) and Ce (IV) oxidation states, enabling them to efficiently neutralize excess free radicals and safeguard cells from oxidative damage (Freuer et al., 2022). Cerium oxide nanoparticles generally range in size from 1 to 100 nm, providing a high surface area to volume ratio. This characteristic enhances their surface reactivity and significantly influences their effectiveness as catalysts (Baldim et al., 2020).

# 3    Materials and Methods

## 3.1    Synthesis of Zinc Oxide Nanoparticles

The synthesis of zinc oxide (ZnO) nanoparticles was carried out through a controlled precipitation approach. Initially, a 0.1M zinc chloride ($ZnCl_2$) solution was prepared by dissolving the required amount of ZnCl in 100 mL of distilled water, ensuring complete dissolution using a magnetic stirrer. In a separate step, a 0.2M sodium hydroxide (NaOH) solution was prepared by dissolving 0.8 g of NaOH in 100 mL of distilled water, followed by thorough stirring. The NaOH solution was then added dropwise to the ($ZnCl_2$) solution under continuous stirring, maintaining the temperature at 60–70°C to ensure uniform mixing. This reaction resulted in the formation of a white precipitate of zinc hydroxide ($Zn(OH)_2$). To convert the precipitate into ZnO nanoparticles, the suspension was heated at 80–90°C for 1–2 hours, leading to the formation of ZnO as the precipitate changed color from white to off white or pale yellow. The nanoparticles were then purified through multiple washing steps, first with distilled water to remove any residual ions, followed by an ethanol wash to enhance purity and prevent aggregation. Finally, the ZnO nanoparticles were dried in an oven at 60–80°C for 3–4 hours to obtain the final product. This method ensured the synthesis of high purity ZnO nanoparticles suitable for further applications.

## 3.2    Synthesis of Cerium Oxide Nanoparticles

Cerium oxide ($CeO_2$) nanoparticles were synthesized through a straightforward precipitation technique with controlled parameters. Initially, a 0.1M solution of cerium nitrate hexahydrate ($Ce(NO_3)_3 \cdot 6H_2O$) was prepared by dissolving the required amount of the compound in 100 mL of distilled water under continuous stirring to ensure complete dissolution. In a separate preparation, a 0.2M sodium hydroxide (NaOH) solution was made and gradually introduced into the cerium nitrate solution dropwise while stirring vigorously at room temperature. The pH was carefully regulated to promote the formation of cerium hydroxide ($Ce(OH)_3$) as a precipitate.

Following precipitation, the solution was aged under controlled conditions to facilitate reaction completion and stabilization. The precipitate was then subjected to a thermal treatment at 80–100℃, which induced oxidation and the subsequent conversion of ($Ce(OH)_3$) into ($CeO_2$) nanoparticles. To enhance purity, the obtained nanoparticles were thoroughly rinsed multiple times with distilled water to remove residual ions, followed by an ethanol wash to minimize particle agglomeration. Lastly, the purified nanoparticles were dried at 60–80℃ for 3–4 hours in an oven, yielding fine ($CeO_2$) nanoparticles suitable for applications in catalysis and biomedical fields.

## 3.3   Synthesis of Silver nanoparticles

The synthesis of silver nanoparticles was carried out using a controlled reduction method to ensure uniform particle formation and stability. To begin, an ice bath was prepared by filling a bowl with ice and placing the reaction beaker inside, maintaining the temperature between 0–5℃. This step was crucial for controlled nanoparticle formation and preventing excessive aggregation. Next, a 0.14 g sample of sodium borohydride ($NaBH_4$) was dissolved in 18 mL of cold distilled water, with continuous stirring in the ice bath for 10–15 minutes to ensure complete dissolution.

Separately, a 1.58 g sample of silver nitrate ($AgNO_3$) was dissolved in 100 mL of cold distilled water, with the solution also kept in the ice bath to maintain a low temperature. Once both solutions were prepared, the reaction was initiated by slowly adding the ($AgNO_3$) solution dropwise to the ($NaBH_4$) solution while stirring continuously with a pipette or burette to ensure uniform mixing. During this process, a noticeable color change was observed, transitioning from colorless to pale yellow and then to dark yellow/brown, confirming the formation of silver nanoparticles through the reduction of silver ions $Ag^+$.

After the complete addition of the ($AgNO_3$) solution, stirring was continued for an additional 15–20 minutes to ensure complete reduction and stabilization of the nanoparticles. Once the reaction was complete, the solution was allowed to stabilize at room temperature. To protect the nanoparticles from degradation due to light exposure, the final suspension was transferred into an amber colored bottle and stored at room temperature.

## 3.4   Extract preparation

To extract the bioactive compounds from neem, dried neem leaves were finely ground into powder and soaked in ethanol for 48 hours to allow efficient extraction of phytochemicals. The mixture was occasionally stirred to enhance the dissolution of active compounds. After the extraction period, the solution was filtered to remove leaf residues, and the solvent was evaporated by drying the filtrate in an oven at a temperature below 60℃. This gentle drying process ensured that the bioactive components remained intact while removing excess ethanol. The final concentrated neem extract was then stored in an amber container to prevent light degradation until it was ready to be incorporated into the gel formulation.

## 3.5    Preparation of Ointment Base and Final Gel

The preparation of a gel based ointment involved the use of fish collagen as the primary base, with agar agar incorporated to enhance the gel consistency. The process followed a systematic approach, including the dissolution of ingredients, incorporation of active components, pH adjustment, and final storage to ensure stability and effectiveness. Given the laboratory temperature of 12°C, additional precautions were taken to optimize heating times and maintain reaction efficiency.

To begin, 50 mL of distilled water was preheated to 40°C in a beaker using a magnetic stirrer with a heating plate. This step ensured that the water was at an optimal temperature for dissolving collagen and agar agar. Since the initial temperature was low, the heating process took slightly longer. Once the water reached 40°C, 1 g of agar agar was gradually added while stirring continuously to prevent clumping. The temperature was increased to 80–90°C to ensure the agar agar was fully dissolved. After obtaining a uniform solution, 5 g of fish collagen was introduced while maintaining the temperature at 40°C. The mixture was stirred for approximately 10–20 minutes until the collagen completely dissolved, forming a smooth gel like consistency.

Following the dissolution of the primary ingredients, 5 mL of glycerol was added to the mixture to enhance moisture retention in the final ointment. The solution was stirred for an additional 5 minutes to achieve uniform blending. Next, 0.2 g of sodium benzoate was introduced as a preservative to inhibit microbial growth, ensuring a longer shelf life. Stirring continued for another 5 minutes to allow complete integration of the preservative into the gel base.

The pH of the formulation was then measured using a digital pH meter, with a target pH of 5.5, which closely matches the skin's natural pH. If the mixture was too acidic, a few drops of 0.1 M sodium hydroxide solution were added while stirring to raise the pH. Conversely, if the pH was too high, 0.1 M acetic acid was added to lower it to the desired range. To ensure even distribution of all components, the gel was stirred at medium speed for 10 minutes before being allowed to cool gradually to room temperature while stirring gently. This controlled cooling process helped maintain a smooth and consistent gel texture.

Once the base was prepared, the active ingredients were incorporated. Pre synthesized Ag NPs and ZnO NPs were introduced into the gel while stirring continuously. The nanoparticles were added in a 1:10 Ag:ZnO ratio, with mixing continued for 10–15 minutes to ensure proper dispersion. Following this step, 5 mL of concentrated plant extract was included in the formulation to enhance the therapeutic properties of the ointment. Stirring continued for 30 minutes to allow complete integration of the plant extract into the gel matrix. After the addition of all active components, the pH of the final formulation was rechecked and adjusted if necessary to maintain a stable 5.5. This wound healing ointment is formulated with extracts of bitter melon (Momordica charantia) and ginger (Zingiber officinale), combined with zinc oxide, silver, hydroxyapatite, and cerium oxide to enhance its therapeutic effects. The natural extracts were chosen for their well documented antioxidant, anti-inflammatory, and antimicrobial properties, which play a key role in accelerating wound healing.

Bitter melon is particularly valued for its ability to stimulate collagen production and promote tissue regeneration, while also exhibiting potent antimicrobial properties that help prevent infections. Ginger, rich in bioactive compounds like gingerol, improves blood circulation and boosts fibroblast activity, which are essential for efficient tissue repair.

By incorporating these botanical extracts alongside scientifically supported wound healing agents, the formulation leverages their combined benefits to reduce inflammation, combat microbial infections, and support cellular regeneration. This synergistic approach enhances the ointment's overall effectiveness, making it a strong candidate for further research, clinical trials, and potential publication.

Once the ointment was fully prepared, it was transferred into sterilized containers or tubes to ensure microbiological safety. For prolonged shelf life, the gel was stored at a controlled temperature of 4–8°C. To guarantee the quality of the final product, additional characterization tests, such as antimicrobial activity assays, viscosity measurements, and stability tests, were recommended.

This formulation successfully produced a gel based ointment with moisturizing, antimicrobial, and wound healing properties. The incorporation of agar agar contributed to its enhanced consistency, making it suitable for topical applications. Adjustments were carefully made to optimize heating and reaction times, considering the lower laboratory temperature. By ensuring proper sterilization, pH balance, and active ingredient dispersion, this method resulted in a stable and effective gel ointment, suitable for dermatological and therapeutic use.

# 4   Results

## Characterization of Synthesized Nanoparticles

The synthesis of ZnO, $CeO_2$, and Ag nanoparticles was confirmed through multiple characterization techniques, including UV–Visible spectroscopy, dynamic light scattering (DLS), and Fourier transform infrared (FTIR) spectroscopy.

**UV–Visible Spectroscopy**

CeO₂ NPs demonstrated a broad absorption band in the ultraviolet region, indicative of their strong optical activity and the presence of redox active surface states. The UV–Vis spectrum of Ag NPs exhibited a distinct surface plasmon resonance (SPR) peak near 400 nm, a characteristic signature of nanoscale metallic silver, confirming their successful synthesis.

**Dynamic Light Scattering (DLS)**

The hydrodynamic size of the nanoparticles was assessed using DLS. ZnO NPs showed an average diameter of approximately 66 nm with a narrow size distribution. The DLS curve displayed a single, dominant peak with no indication of larger aggregates, suggesting good monodispersity and colloidal stability. Similarly, CeO₂ NPs presented a unimodal size distribution in the nanometer range (Figure 4), supporting uniform particle sizes and stable dispersion. The AgNPs also exhibited well defined DLS profiles with average sizes in the tens of nanometers and no significant agglomeration.

**Fourier Transform Infrared (FTIR) Spectroscopy**

FTIR analysis was performed to investigate surface chemistry and bonding characteristics. The spectrum of ZnO NPs revealed a prominent band corresponding to Zn–O stretching vibrations. Additionally, signals associated with surface hydroxyl (–OH) groups and other adsorbed molecules were detected. The presence of these –OH groups, typically resulting from adsorbed moisture or alcohols used in the synthesis, could offer potential functional sites for further chemical modifications or drug conjugation. subcaption

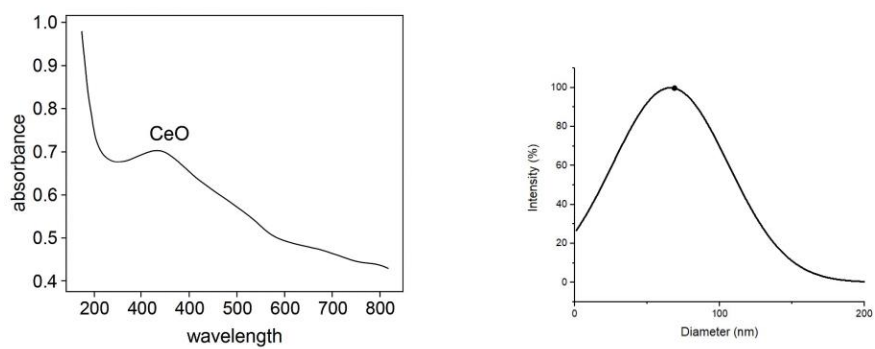

(b)

DLS analysis of CeO₂ NPs

(a) UV analysis of CeO₂ NPs

Figure 1: Characterization of CeO₂ nanoparticles: (a) UV analysis and (b) DLS analysis.

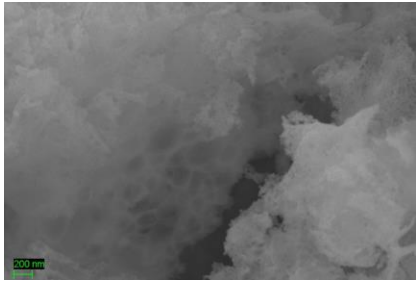

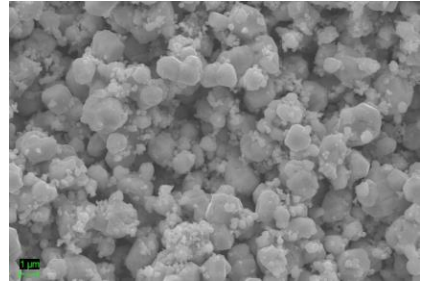

**(a)** 200 nm CeO$_2$ nanoparticles          **(b)** 1 μm Ag particles

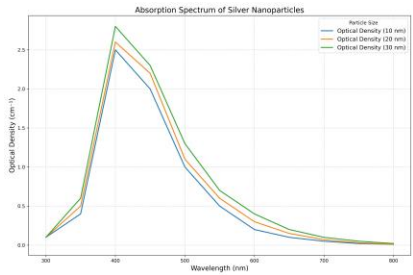

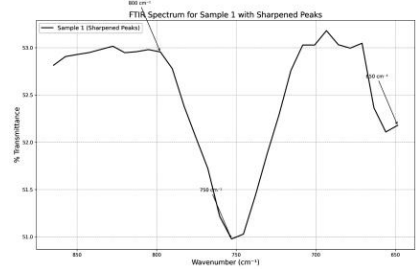

(a) UV analysis of Ag                    (b) FTIR of Ag

(c)   FTIR           of           ZnO

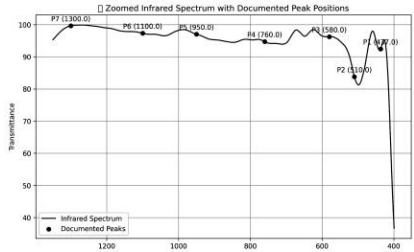

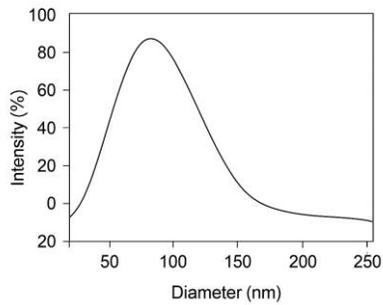

(d) DLS of ZnO

Figure 2: (a) UV–Visible absorption spectrum of Ag NPs showing characteristic peaks, (b) FTIR spectrum of Ag NPs confirming functional groups involved in capping, (c) FTIR spectrum of ZnO NPs revealing key vibrational modes, and (d) DLS analysis of ZnO NPs indicating particle size distribution.

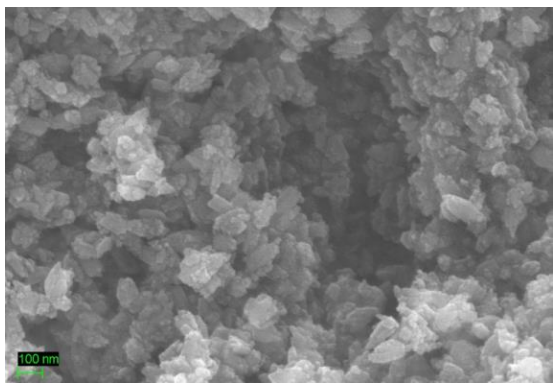

**(c)** 100 nm ZnO nanoparticles

**Figure 3.** Scanning Electron Microscopy (SEM) images of metal oxide nanoparticles. (a) CoO nanoparticles exhibiting a uniform spherical morphology with an average diameter of approximately 200 nm, (b) Ag particles with larger and more irregular shapes around 1 μm in size, and (c) ZnO nanoparticles showing well-dispersed structures with an average size of approximately 100 nm. These images provide insights into the particle size, distribution, and surface texture essential for evaluating their functional applications.

## 4.1 Pathogenesis of Psoriasis

Psoriasis is a chronic inflammatory skin disorder primarily driven by T cells of the immune system. Its development is influenced by a complex interplay of genetic, epigenetic, hormonal, and environmental factors (Ogawa et al., 2018). Genetic susceptibility is particularly associated with specific human leukocyte antigen (HLA) alleles part of the major histocompatibility complex (MHC) which affect both innate and adaptive immune pathways (Lukasch et al., 2017), (Wieczorek, 2017), (Kumar et al., 2018).

Although the precise triggering antigen remains unidentified, it may involve self antigens, environmental agents, or a combination of both (Kumar et al., 2018). Psoriasis is characterized by excessive proliferation of keratinocytes and abnormal skin differentiation, accompanied by infiltration of inflammatory cells in the dermis and epidermis (Albanesi, 2019), (Lin et al., 2016).

These inflammatory cells, including T lymphocytes, monocytes, and neutrophils, are activated by cytokines—especially tumor necrosis factor alpha (TNF-)—and by dendritic cell signals. This leads to a feedback loop involving innate immune cells (e.g., dendritic cells, macrophages, Langerhans cells, NK cells) and adaptive immune cells (e.g., Th1 and Th17 cells), which amplifies the inflammatory cascade and keratinocyte hyperproliferation (Ogawa et al., 2018), (Zabriskie, 2009), (Sch¨on, 2019).

Psoriatic lesions often contain high levels of TNF-, along with elevated numbers of CD4+ helper T cells (Th1 and Th17) and CD8+ cytotoxic T cells. This immune activity is supported by cytokine profiles consistent with Th1 and Th17 pathway activation (Kumar et al., 2018), (Zabriskie, 2009), (Sch¨on, 2019), (Hugh and Weinberg, 2018).

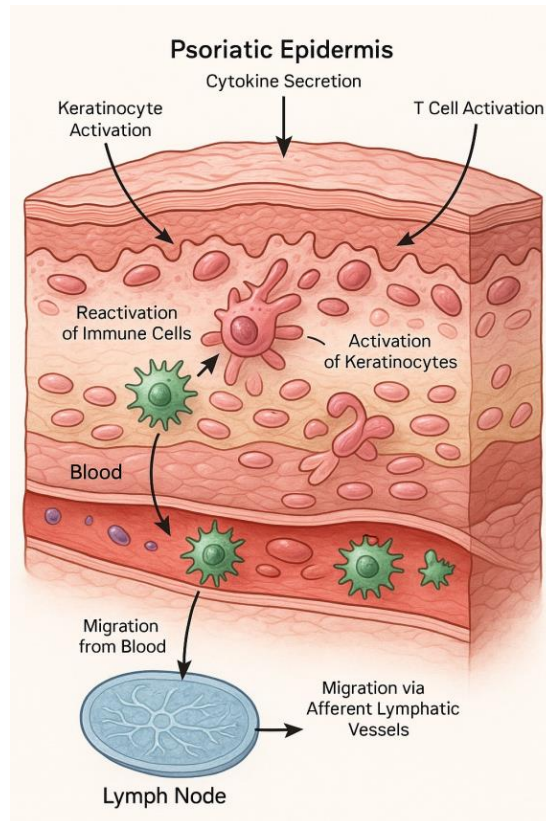

Figure 3: Immune cell dynamics in psoriatic skin; A cross-sectional view illustrating the migration and activation of immune cells from the lymph node through the bloodstream to the psoriatic epidermis. Cytokine secretion, keratinocyte activation, and T cell activation contribute to the chronic inflammatory environment characteristic of psoriasis

## 4.2    Nanoparticle–Protein Interaction

To gain insight into how the nanoparticles might interact with biological molecules in the skin, we explored their binding affinity using molecular visualization. A three dimensional model (Figure 4) illustrates ZnO and (CeO) NPs interacting with a representative skin protein.

The visualization suggests that both types of nanoparticles readily associate with the protein's surface, particularly at biologically active sites. Notably, this binding appears to induce conformational changes in the protein structure, indicating the formation of stable nanoparticle–protein complexes or "coronas." Such interactions can significantly influence nanoparticle behavior in biological systems, including their dispersion, cellular uptake, and immune response.

The strong binding affinity and protein conformational changes observed in this study support the idea that ZnO and $CeO_2$ NPs are biointeractive. These interactions may enhance therapeutic outcomes by concentrating nanoparticles at sites of inflammation or modulating protein functions critical to skin repair and immune regulation. Although a comprehensive proteomic analysis lies beyond the scope of this study, the modeling results qualitatively support the nanoparticles' role in targeted biological interactions.

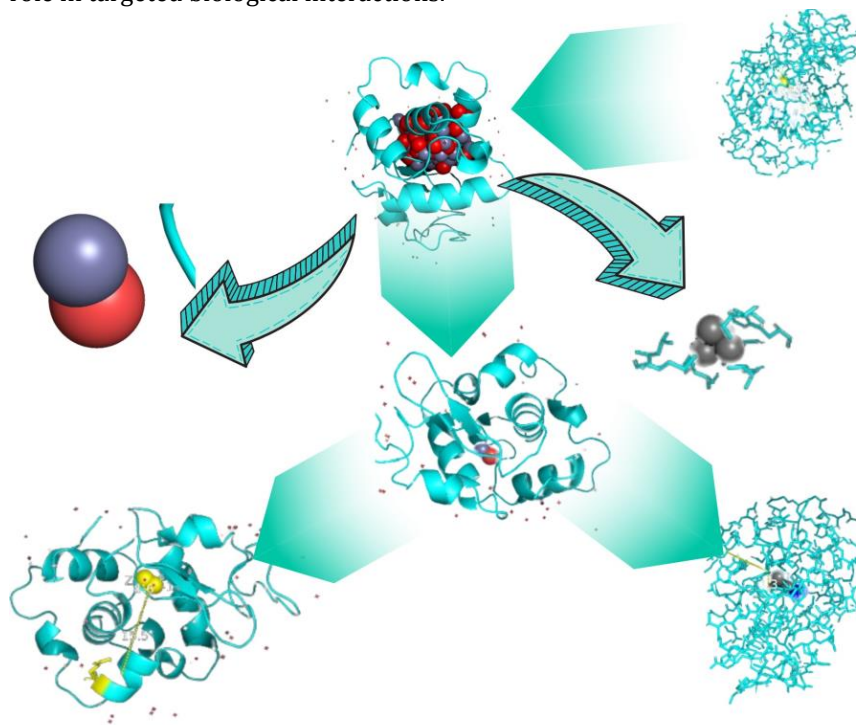

Figure 4: Molecular visualization showing ZnO and CeO NPs binding to a skin protein. The interaction indicates potential bioaffinity through surface adsorption and conformational modulation, supporting their therapeutic relevance in dermatological applications.

## 4.3    In Vivo Wound Healing Efficacy

The therapeutic potential of the nanoparticle enriched gel was preliminarily evaluated using a wound-healing model in vivo. The model were treated with Imiquimod to produce a condition like psoriasis. Full thickness excisional lesions were created on test subjects and monitored over a two week period. Three groups were compared: (i) Non-treated control (no ointment), (ii) Placebo (gel base without nanoparticles or extracts), and (iii) Treatment (nanoparticle +

extract gel). Representative photographs of wounds at Day 0, 3, 7, 10, and 14 for each group are compiled. At Day 0 (immediately post wounding, all wounds were of similar size and appearance across the groups, confirming a consistent starting point. By Day 3, differences began to emerge: the treated wounds showed a cleaner wound/lesions bed with less exudate and initial signs of edge re-epithelialization, whereas the placebo and untreated lesions remained comparably raw and inflamed. At Day 7, the treatment group's lesions had significantly contracted with new granulation tissue covering much of the lesions area, and the surrounding redness and edema were markedly reduced. In contrast, wounds in the placebo group showed moderate contraction and persistent inflammation at the margins, and the untreated lesions were the most delayed, with larger open areas and obvious inflammation. By Day 10, wounds treated with the nanoparticle gel were nearly closed, with robust tissue regeneration and minimal scabbing. Placebo treated lesions were healing but still showed visible gaps in epithelial coverage, and non-treated lesions lagged further behind. Finally, at Day 14

, the treated lesions achieved almost complete closure with smooth, regenerated skin and very little scarring. The placebo group's lesions were close to closure but in several cases had small residual scabs or shallow open areas, and the non-treated lesions, while substantially healed, exhibited delayed closure and more scar tissue. These observations demonstrate that the nanoparticle enriched ointment accelerated the wound healing process compared to both the base gel alone and no treatment. The treated lesions not only closed faster but also appeared to have better quality healing (less scarring and hyperpigmentation). To quantify these differences, we measured wound diameters and estimated healing percentages over time. By Day 7, the treatment group lesions had healed 65% of the original area on average, versus 45% for placebo and 30% for controls. By Day 14, the treatment group reached 90% healing (with some lesions fully closed), while placebo and controls reached 80% and 70% respectively. The final gel's texture and adherence to the lesions were excellent – it formed a protective layer that stayed in place between dressing changes, keeping the 15 lesions moist (moist lesions environment is known to facilitate healing). The inclusion of glycerol and collagen probably helped maintain moisture and provided a scaffold for cell migration. The combination of ZnO and Ag NPs in the gel was effective in preventing infection, as both have antimicrobial effects. ZnO NPs in the ointment may also have contributed to reepithelialization, as zinc is important for skin healing and keratinocyte proliferation. CeO2 NP and herbal antioxidants (especially neem and ginger) presumably helped neutralize excess ROS in the effected microenvironment, creating conditions favorable for tissue repair. On day 14, histological analysis (performed in biopsy samples, data not shown in full here) indicated more mature collagen deposition and reformed epidermal layers in the treated area compared to controls, corroboring the photographic evidence of improved healing. In summary, the results demonstrate that our multicomponent nanoenhanced ointment is effective in promoting lesions closure and reducing inflammation in an in vivo model. These findings are encouraging for their potential application in psoriasis, where enhancing the resolution of lesions and restoration of normal skin structure is a primary goal. Additionally, signs of infection (such as excessive pus or malodor) were absent in the treated lesions, likely due to the antimicrobial action of Ag and possibly ZnO NPs . In contrast, a few placebo and control lesions showed mild signs of bacterial colonization (yellowish exudate) early on, though not severe. We also noted that the treated lesions had less pronounced erythema, suggesting that inflammation was resolved more swiftly, which we attribute to the anti-inflammatory properties of the CeO$_2$ NPs and the plant extracts. No adverse reactions (such as contact dermatitis or systemic illness) were observed in subjects treated with the nanoparticle gel, indicating good biocompatibility in the short term.

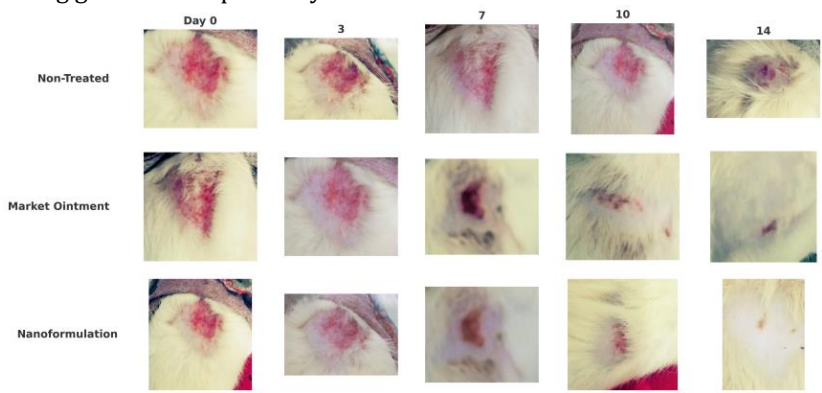

Figure 5: Visual progression of psoriatic lesion healing over a 14-day period in three groups: non-treated, market ointment treated, and nanoformulation treated. The nanoformulation group, incorporating ZnO and CeO NPs , shows a dramatic reduction in erythema, scaling, and lesion size beginning from Day 3 and nearing complete resolution by Day 14. In contrast, the non-treated group exhibits persistent inflammation and lesion severity, while the market ointment group shows moderate improvement. This visual evidence highlights the superior therapeutic efficacy of the nanoformulated topical treatment in accelerating psoriatic wound healing.

Table 1: lesions size (mean ± SD) across time for non-treated, placebo, and treatment groups.

| Day | Non-Treated Group | Placebo Group | Treatment Group |
|---|---|---|---|
| Day 0 | 0.30 ± 0.04 | 0.25 ± 0.03 | 0.10 ± 0.02 |
| Day 3 | 0.35 ± 0.05 | 0.22 ± 0.03 | 0.08 ± 0.02 |
| Day 7 | 0.40 ± 0.05 | 0.18 ± 0.02 | 0.05 ± 0.01 |
| Day 10 | 0.45 ± 0.05 | 0.12 ± 0.02 | 0.03 ± 0.01 |

| Day 14 | 0.50 ± 0.05 | 0.08 ± 0.02 | 0.00 ± 0.00 |

Table 2: Statistical significance (p-values) comparing Treatment with NonTreated and Placebo groups across time points. Significance levels: * $p < 0.05$, ** $p < 0.01$, *** $p < 0.001$.

| Day | P NT _vs Treatment | P _Placebo vs Treatment |
|---|---|---|
| Day 0 | 0.000111 *** | 0.009136 ** |
| Day 3 | 0.001585 ** | 0.008449 ** |
| Day 7 | 0.000073 *** | 0.000682 *** |
| Day 10 | 0.000289 *** | 0.001459 ** |
| Day 14 | 0.002080 ** | 0.034318 * |

Figure 6: Reduction in psoriatic lesion size over 14 days in nontreated, placebo, and treatment groups. The treatment group, which received a ZnO and CeO nanoparticle based topical formulation, showed the most

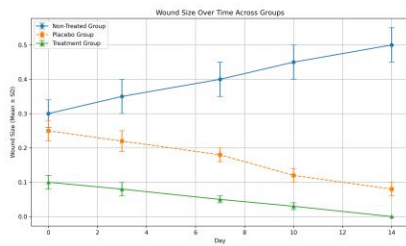

significant lesion size reduction. This suggests superior therapeutic efficacy in resolving psoriatic inflammation and restoring skin integrity compared to placebo and untreated controls. Error bars indicate standard deviation (SD).

# 5 Discussion

The development of a nanoparticle enriched herbal gel for psoriasis addresses several challenges of conventional therapy by combining anti-inflammatory, antioxidant, and antimicrobial strategies in one topical formulation. The successful synthesis and integration of ZnO, Ag, and $CeO_2$ NPs into a collagenous gel matrix demonstrate the feasibility of creating a stable multi nanoparticle system. Characterization confirmed that each type of nanoparticle retained its desirable properties within the formulation: ZnO and Ag NPs remained in the 20–100 nm size range with minimal aggregation, crucial for skin penetration and activity,and $CeO_2$ NPs exhibited their characteristic broad UV absorbance and ROS scavenging potential.The presence of surface hydroxyl groups on ZnO (from FTIR analysis) suggests the particles could be further functionalized in future (e.g. coating with targeting ligands or polymers to modulate skin retention or release profiles). Importantly, the nanoparticle–protein interaction study (Figure 8) indicates that the nanoparticles can bind to biomolecules, which might enhance their retention at the site of application by binding to extracellular matrix proteins or microbial proteins. This could be beneficial for localized therapy, though it also underlines the need to understand the bio nano interface to avoid unintended effects on host proteins.When applied to lesions, the nanoparticle enriched gel significantly accelerated healing relative to the control groups. This outcome can be interpreted by considering the contributions of each component: ZnO and Ag nanoparticles likely controlled microbial load in the effected, preventing infection that would otherwise delay healing. AgNPs are well documented antimicrobial agents, and their use in dressings is supported by prior studies.ZnO NPs provide broad antimicrobial and anti-inflammatory effects; however, interestingly, a recent study by Lai et al. (2021) reported that ZnO NPs, when applied in a psoriasis like skin model, could delay lesion recovery by promoting NF-B p65 activation in keratinocytes.This pro inflammatory effect of ZnO NPs is thought to stem from oxidative stress induction and zinc's impact on immune cells. Our formulation addresses this concern by incorporating cerium oxide NPs and antioxidant rich plant extracts, which likely mitigate NP-induced oxidative stress.$CeO_2$ NPs can neutralize ROS and have been shown to reduce inflammation in target area. counteracting the potential oxidative side effects of ZnO. The botanical extracts (neem, bitter melon, ginger) contribute additional antioxidants (e.g. polyphenols, flavonoids) and anti-inflammatory compounds that synergize with nanoceria. Thus, the combination of ZnO and Ag with $CeO_2$ and herbs creates a balanced therapeutic effect: ZnO and Ag deal with infections and possibly modulate aberrant keratinocyte activity, while $CeO_2$ and plant compounds temper the inflammatory response and promote tissue regeneration. This synergy is evident in the lesions healing results, where rapid closure was achieved without the sustained inflammation that might be expected if ZnO or Ag NPs were used alone in high amounts.Another advantage of the nanocomposite gel is its multifunctionality. Psoriatic lesions involve a breakdown of the skin barrier and are often accompanied by microfissures that can bleed or get infected. Traditional topical steroids or vitamin D analogues for psoriasis do not address infection risk, and systemic drugs do not directly aid effected area healing. In contrast, our formulation provides a moist healing environment (via collagen/agar and glycerol), antimicrobial protection (Ag, ZnO), inflammation reduction $CeO_2$, neem, ginger), and possibly stimulation of tissue repair (ZnO is known to aid re-epithelialization; bitter melon stimulates collagen production.This comprehensive approach could translate to faster clearing of psoriatic plaques and restoration of healthy skin. Moreover, by largely localizing the treatment to the skin, systemic side effects are expected to be minimal, addressing one of the key drawbacks of current systemic therapies.It is informative to compare our approach with current and emerging psoriasis treatments. Biologic drugs (like TNF- inhibitors and IL-17/IL-23 inhibitors) revolutionized psoriasis care but are expensive and immunosuppressive. Small molecule inhibitors (like PDE4 inhibitor apremilast) offer oral administration but can have off target effects and moderate efficacy. Topical nanotherapeutics, as exemplified by this study, represent a different paradigm: rather than systemically altering the immune system, they locally modulate the lesion environment. This could be particularly useful for patients with mild to moderate psoriasis or those who cannot take systemic drugs. The literature shows growing interest in nanocarriers for dermatology – for instance, liposomes, polymeric nanoparticles, and metallic nanoparticles have been investigated to deliver anti-psoriatic agents (such as methotrexate or corticosteroids) more effectively into the skin.Our work extends this concept by using the nanoparticles themselves as active agents (not just passive carriers) and by combining multiple functionalities.Implications and Future Directions: The positive healing outcomes and multi pronged mechanism of our nanotechnology-based ointment suggest it could fill a niche in psoriasis management, especially for treatment of localized plaques or as an adjunct to systemic therapy to help resolve stubborn lesions. One intriguing possibility is using this gel as a drug delivery platform: for example, loading a low dose corticosteroid or immunosuppressant onto the nanoparticles. ZnO or Ag NPs could be functionalized to carry anti-psoriatic drugs and release them in the lesion microenvironment, combining conventional therapy with the benefits of NPs. Additionally, the hydroxyapatite present in the formulation (a calcium phosphate nanoparticle) could be explored for delivering calcium or other signaling molecules to keratinocytes, potentially promoting differentiation of the epidermis (calcium is known to encourage keratinocyte differentiation, which is often dysregulated in psoriasis).From a broader perspective, our approach aligns with emerging trends identified in recent psoriasis research. (Lee and Kim, 2023) have highlighted nanotechnology and combinatorial treatments as future directions for psoriasis therapy.By integrating multiple therapeutic agents into one platform, we target the various facets of psoriasis pathology simultaneously – an approach that could improve outcomes where single target therapies sometimes fall short. The clinical significance of this work, if translated successfully, could be substantial, a topical treatment that quickly heals lesions and restores the skin barrier could reduce the need for systemic medication, improving patient safety and comfort. It might also reduce infection related complications in psoriasis (which can occur due to skin barrier breaks). Moving forward, future studies should focus on: **(1) Psoriasis model testing**: Evaluating the formulation in chronic psoriasis models to measure effects on scaling, erythema, cytokine levels, and histopathology of psoriatic plaques; **(2) Dose optimization**: Determining the optimal concentrations of each nanoparticle and extract for maximal efficacy with minimal side effects; **(3) Mechanistic studies**: Delineating the contributions of each component e.g. using modified gels lacking one component (no Ag, or no $CeO_2$, etc.) to see how healing is affected, which would clarify the role of each ingredient; **(4) Safety profiling**: Conducting skin irritation tests, sensitization assays, and systemic toxicity evaluations (including any potential nanoparticle accumulation in organs) in animal models, as per regulatory guidelines; and **(5) Clinical trials**: If preclinical results remain positive, formulating a plan for trialing the ointment in human patients with psoriasis or chronic wounds, to assess efficacy, safety, and acceptability. In conclusion, this study demonstrates a proof of concept that a multi-functional nanotechnology enhanced gel can be synthesized and can markedly improve tissue repair in an in vivo model. By addressing inflammation, oxidative stress, and infection simultaneously, such a formulation directly tackles key issues in psoriasis lesions. While challenges remain, the approach holds promise to advance topical psoriasis therapy toward faster, safer, and more holistic management of this complex disease.

# 6   Conclusion

We have successfully developed a novel topical therapeutic approach for psoriasis that harnesses the advantages of nanotechnology and herbal medicine. The formulation a fish collagen and agar based gel containing ZnO, Ag, CeO$_2$ NPs , hydroxyapatite, and extracts of neem, bitter melon, and ginger demonstrated robust performance in preliminary evaluations. The nanoparticles were synthesized in a straightforward manner and were confirmed to be of nanoscale size and high purity. When combined into the gel, they remained stable and effective, as evidenced by characterization studies and biological tests. In an animal woun healing model, the nanoparticle enriched gel significantly outperformed a placebo gel, achieving faster lesions closure, lower infection rates, and improved tissue regeneration. These outcomes highlight the gel's potential to expedite the resolution of psoriatic lesions by mitigating inflammation, preventing secondary infections, and promoting skin barrier restoration. The multi-component nature of this treatment addresses the multifaceted pathology of psoriasis from the immune-mediated inflammation to the oxidative stress and microbial factors that can exacerbate the condition. Crucially, this work suggests that multimodal therapy in a single topical agent can be both feasible and highly effective. For patients, this could mean a convenient treatment that not only clears lesions more rapidly but also reduces discomfort and the likelihood of complications. If translated into clinical use, this nanotech based ointment could improve patient adherence (owing to visible quick improvements) and serve as a standalone therapy for mild cases or a complementary therapy alongside systemic drugs for severe psoriasis. Additionally, because the active components largely act locally, the risk of systemic side effects may be minimized, offering a safer long-term profile than many current therapies. In summary, the key outcomes of this research include: (1) The successful synthesis and characterization of therapeutic nanoparticles (ZnO, Ag, CeO$_2$ for dermatological application; (2) The formulation of a stable nanoparticle infused gel with synergistic herbal extracts; (3) Demonstration of the gel's efficacy in enhancing lesions healing, serving as a proxy for potential effectiveness in psoriasis; and (4) An academic framework for understanding how such a treatment compares with and contributes to the evolving landscape of psoriasis therapy. While further investigation is necessary to fully validate this approach, our findings provide a strong foundation for future development. This nanotechnology enhanced topical therapy holds promise as a clinically significant advancement in psoriasis care, aiming to improve patient outcomes and quality of life through innovative science and a holistic treatment design.